\documentclass[aps,twocolumn,prb,superscriptaddress]{revtex4-2}
\usepackage[utf8]{inputenc}

\usepackage{natbib}
\usepackage{graphicx}
\usepackage{dcolumn}
\usepackage{bm}
\usepackage{epsfig}
\usepackage{textcomp}
\usepackage{gensymb}
\usepackage{xcolor}
\usepackage{placeins}
\usepackage{physics}
\usepackage{amssymb,amsmath,amsthm,enumitem}
\usepackage{float}

\usepackage{comment}
\usepackage{caption}
\usepackage{subcaption}
\usepackage{booktabs}
\usepackage{lipsum}
\usepackage{placeins}
\usepackage[version=4]{mhchem}
\usepackage{hyperref}
\usepackage{dcolumn}
\usepackage{ragged2e}
\usepackage[normalem]{ulem}
\usepackage{multirow}

\begin{document}
\title{Excitation spectrum and low-temperature magnetism in disordered defect-fluorite 
\ce{Ho2Zr2O7}}
\author{P. L. Oliveira Silva}
\email{pedrolucas@usp.br}
\affiliation{Instituto de Física, Universidade de São Paulo, 05314-970 São Paulo, Brazil}

\author{J.G.A. Ramon}
\affiliation{Jülich Centre for Neutron Science (JCNS), Heinz Maier-Leibnitz Zentrum (MLZ), Forschungszentrum Jülich GmbH, Lichtenbergstr. 1, D-85747 Garching, Germany}
\affiliation{Instituto de Física, Universidade de São Paulo, 05314-970 São Paulo, Brazil}

\author{Viviane Peçanha-Antonio}
\email{viviane.antonio@stfc.ac.uk}
\affiliation{ISIS Facility, Rutherford Appleton Laboratory, Chilton, Didcot OX11 0QX, United Kingdom}
\affiliation{Department of Physics, University of Oxford, Clarendon Laboratory, Oxford OX1 3PU, United Kingdom}

\author{Tatiana Guidi}
\affiliation{School of Science and Technology, Physics Division, University of Camerino, I-62032 Camerino, Italy}
\affiliation{ISIS Facility, Rutherford Appleton Laboratory, Chilton, Didcot OX11 0QX, United Kingdom}

\author{J. S. Gardner}
\affiliation{Oak Ridge National Laboratory, Oak Ridge, Tennessee 37831, USA}

\author{Chun Sheng Fang}
\affiliation{Songshan Lake Materials Laboratory, Dongguan, Guangdong 523808, China}

\author{R. S. Freitas}
\email{freitas@if.usp.br}
\affiliation{Instituto de Física, Universidade de São Paulo, 05314-970 São Paulo, Brazil}

\begin{abstract}

In this work, we report on the thermomagnetic characterization and crystalline-electric field (CEF) energy scheme of the disordered defect-fluorite \ce{Ho2Zr2O7}. This structural phase is distinguished by the coexistence of magnetic frustration and extensive disorder, with \ce{Ho^{3+}} and \ce{Zr^{4+}} sharing randomly the same $4a$ site with even 50\% occupancy, and an average 1/8 oxygen vacancy per unit cell. AC magnetic susceptibility measurements performed on powder samples down to 0.5\,K revealed signs of slowing spin dynamics without glassy behavior, including a frequency dependent peak at $\sim1$\,K. Yet, no evidence for long-range magnetic order is found down to 150\,mK in specific heat. Inelastic neutron scattering measurements show a weak, low-lying CEF excitation around $2$\,meV, accompanied by a broad level centered at $60$\,meV. To fit our observations, we propose an approach to account for structural disorder in the crystal-field splitting of the non-Kramers \ce{Ho^{3+}}. Our model provides an explanation to the broadening of the high-energy, single-ion excitations and suggests that the zirconate ground-state wave function has \emph{zero magnetic moment}. However, structural disorder acts as guarantor of the magnetism in \ce{Ho2Zr2O7}, allowing the mixing of low lying states at finite temperatures. Finally, we show that this scenario is in good agreement with the bulk properties reported in this work.
\end{abstract}

\maketitle

\section{INTRODUCTION}

A system in which all magnetic interactions cannot be simultaneously satisfied is said to be magnetically frustrated. The presence of frustration typically precludes the formation of ordinary ground states and instead favors the emergence of strongly correlated behavior of various kinds \cite{votja1}. Rare-earth pyrochlores are three-dimensional frustrated systems with a long record of exotic magnetic phenomena \cite{gardner2010magnetic,rau1,roll1,Scheie1,ramirez1999zero} and have been at the forefront of the search for quantum spin liquid (QSL) phases \cite{poree1,savary1,sibille1,gingras1,sibille2}. 

In pyrochlores with formula \ce{\it{A}2\it{B}2O7}, the rare-earth \ce{\it{A}^{3+}} and transition-metal \ce{\it{B}^{4+}} ions form two independent corner-sharing tetrahedral sublattices \cite{subramanian1}. The local symmetry $D_{3d}$ of the CEF environment, combined with strong spin-orbit coupling, plays a crucial role in the magnetic behaviour of the rare-earth ions. A canonical example of that is the “2-in-2-out” ice rule of the dipolar spin ice \ce{Ho2Ti2O7} \cite{harris1997geometrical,bramwell2001spin,den2000dipolar}. In this compound, the non-Kramers \ce{Ho^{3+}} exhibits a doublet ground state, separated by the crystalline-electric field by a 200\,K gap from the first excited state. This strong anisotropy restricts the spins to align along the local $\langle111\rangle$ axis, leading to an extensive degeneracy \cite{henley2010coulomb} and magnetic monopole excitations \cite{castelnovo2008magnetic,jaubert2011magnetic}.

Notably, many of these pyrochlores compounds exhibit at least some degree of charge disorder (e.g., holes or doping) \cite{takatsu2016quadrupole,kadowaki2018continuum,taniguchi2013long}. Theoretical models suggest disorder may lift up ground-state degeneracy and promote spin freezing \cite{saunders2007spin,shinaoka2011spin,sen2015topological} or increase the systems’ degrees of freedom, enhancing competition between different states. Especially for non-Kramers ions, that may favor quantum entanglement \cite{savary2017disorder}. It was suggested in Alexanian \emph{et al.} \cite{alexanian2023collective}, that charge disorder at the \emph{B} site acts as a tuning parameter towards a strongly-correlated magnetic ground state, while in Porée \emph{et al.} \cite{poree2022crystal}, \ce{Ce^{4+}} defects are believed to significantly alter CEF excitations. In Sibille \emph{et al.} \cite{sibille2017coulomb}, anion Frenkel disorder seems to induce a dynamical Coulomb spin-liquid phase.

In contrast with the relatively small amount of disorder existing  in pyrochlores, extensive disorder is present in some members of the \ce{\it{A}2\it{B}2O7} family -- the defect, or disordered fluorites. This crystalline phase becomes energetically favorable when the ionic ratio $r_A/r_B$ is less than 1.46 \cite{reynolds2013anion}. In these systems, \ce{\it{A}^{3+}} and \ce{\it{B}^{4+}} share the same lattice site with 50\% of occupancy each, while charge neutrality is maintained by an average 1/8 oxygen vacancy per unit cell \cite{lumpkin2021perspectives}. 

Previous bulk experiments on the defect fluorites \ce{Ho2Zr2O7} \cite{sheetal2022field,elghandour2024slow} and \ce{Dy2Zr2O7} \cite{ramon2019absence,devi2024muon} showed ground states distinct from their pyrochlore counterparts, \ce{Ho2Ti2O7} and \ce{Dy2Ti2O7}. Magnetic susceptibility and muon spin relaxation measurements down to 50\,mK in \ce{Dy2Zr2O7} \cite{ramon2019absence} suggest a disorder-induced dynamical ground state. Similarly, spin freezing below 0.6\,K \cite{elghandour2024slow} is suggested by AC susceptibility studies on \ce{Ho2Zr2O7}, without signs of long-range order down to 0.28\,K. On the other hand, CEF excitations in fluorites remain experimentally unexplored. As previously discussed for the pyrochlores, they are fundamental to understand the magnetic ground state. However, despite the highly symmetric $O_h$ point-group of the magnetic site, site mixing on the $4a$ Wyckoff position and the random oxygen vacancies in fluorites pose a significant challenge to the data analysis of single-ion excitations.

In this work, we investigate the disordered fluorite \ce{Ho2Zr2O7} by measuring its thermomagnetic bulk response down to mK temperatures and characterizing its CEF splitting via neutron spectroscopy. To resolve the CEF scheme of the non-Kramers \ce{Ho^{3+}} ion, we employ two distinct models. One, which we shall refer to as \emph{standard model} (S), is constructed assuming an $O_h$ local point symmetry for the magnetic site, and the other, the \emph{effective model} (E) is built in an attempt to quantify how less-symmetric environments influence the observations. Our results indicate that, for the S model, the ground-state wavefunction is a non-magnetic doublet whereas, for the E model, the ground-state wavefunction is a singlet. Both models have small gaps of $\sim0.9$\,meV to the first excited state. Our analysis suggests that the presence of close, low-lying energy levels is favored by the low-symmetry environment in this compound. Moreover, we show that this characteristic might explain, at least in part, the intrinsic broadening observed in the higher excited modes.

\section{METHODS}

Two different synthesis methods, sol-gel reaction and solid state, were used to prepare the samples used in this work. For the sol-gel reaction, the stoichiometric reagents were prepared as two solutions, one with \ce{Ho2O3}, \ce{HNO3}, and a few drops of deionized water, and other with \ce{Zr(OC4H9)4} and \ce{C6H8O7} diluted in \ce{C2H5OH}. The solutions were continuously stirred at 80°C until a gel was formed and calcined at 950°C for 24 hours, resulting in a yellow powder. For the solid state reaction, stoichiometric quantities of \ce{Ho2O3} and \ce{ZrO2}, were mixed and heated at 1100ºC for 48h. This process was repeated until only the fluorite phase could be detected in the X-ray diffraction pattern.

X-ray diffraction was performed on both samples on a Bruker AXS Discover diffractometer with Cu-K$\alpha_1$ ($\lambda=1.5406\,\text{\AA}$) radiation in Bragg-Brentano geometry. Magnetic measurements were carried out on the sol-gel sample on a superconducting quantum interference device (SQUID, Quantum Design) and a home-built AC susceptometer insert equipped with a $^3$He cryostat, operating at a 155\,Hz frequency and 10\,Oe field down to 0.5\,K. AC magnetic susceptibility was performed down to 50\,mK on a demagnetization refrigerator (Cambridge Cryogenics) in a frequency range of 2\,Hz to 10\,kHz. Specific heat studies were conducted on the calorimeter module of a PPMS (Quantum Design) equipped with a dilution refrigerator reaching temperatures down to 50\,mK. Addenda (Apiezon Grease and platform contributions) were measured prior to the experiment and subtracted from the final data.

Inelastic neutron scattering data were collected on the MARI spectrometer \cite{MARI} at the ISIS Neutron and Muon Source. The \ce{Ho2Zr2O7} sample utilized in these experiments, synthesized via the solid-state reaction, was wrapped in aluminium foil and loaded in a standard Al can, before being placed in the instrument's closed-cycle refrigerator (CCR). Data were recorded at 7, 100 and  200 K using repetition-rate multiplication (RRM) mode, which enabled simultaneous measurements with incident neutron energies $E_\mathrm{i}=10,{ }50$ and 120\,meV.

\section{RESULTS}

\subsection{Powder X-ray diffraction}

\begin{figure}
    \centering   
    \includegraphics[width=\linewidth]{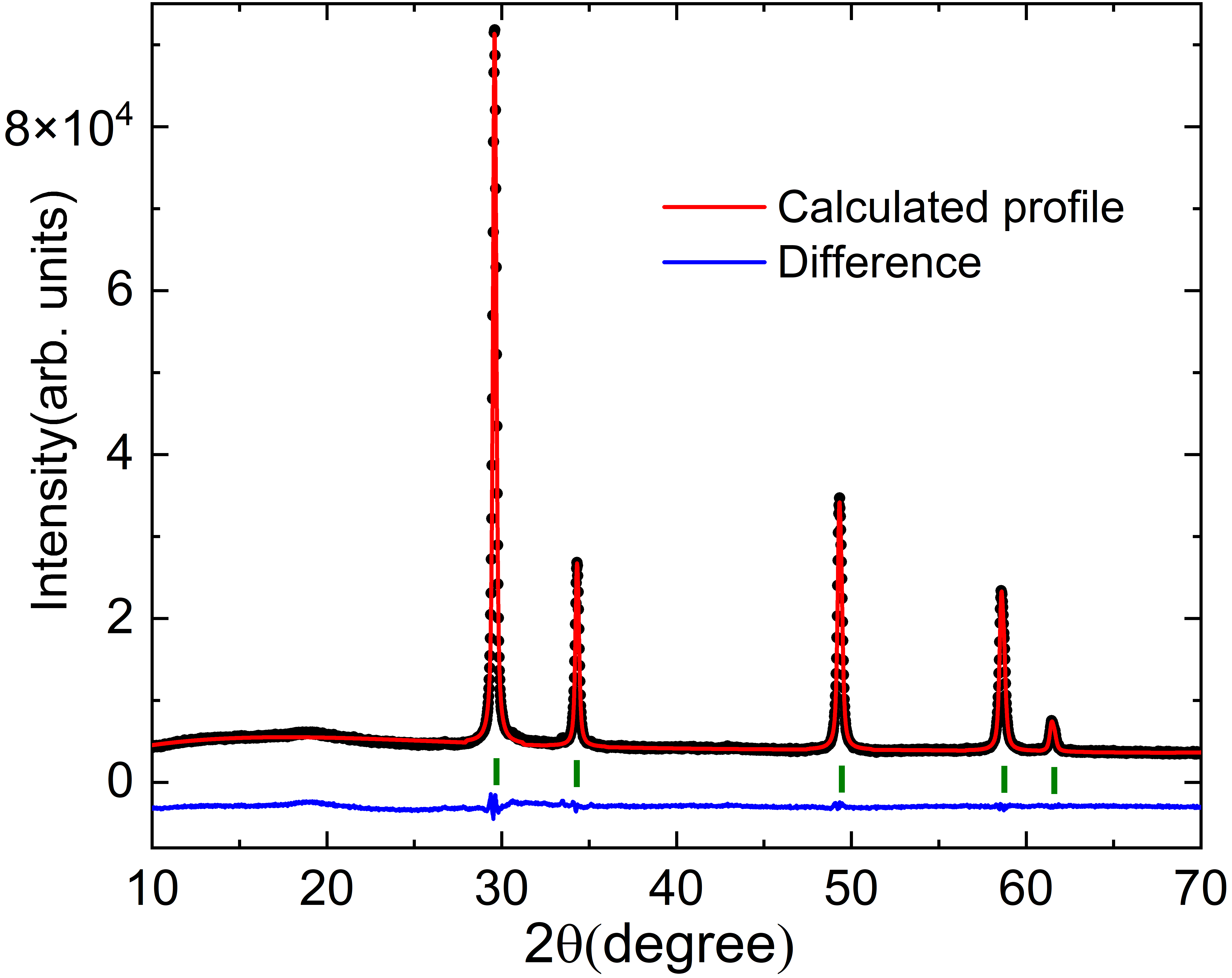}
    \caption{\justifying X-ray diffraction pattern of sol-gel synthesized \ce{Ho2Zr2O7} sample (black circles), showed along with the refined defect-fluorite phase (red line). Difference between data and refinement is shown at the bottom of the figure (blue line).}
    \label{fig:XRD}
\end{figure}

Figure \ref{fig:XRD} shows powder X-ray diffraction data collected on the sample synthesized via the sol-gel reaction. Structure refinement was performed using the Fullprof software suite \cite{rodriguez1993recent}.
No signal of additional phase or impurity was found in the sol-gel nor in the solid-state reaction samples. We refined the data as a defect-fluorite structure with the space group $Fm\bar{3}m$. Calculation (red curve) is consistent with experiment, and had a $\chi^2=7.2$ goodness-of-fit. The refined lattice parameter $a=5.216(2)\text{\AA}$ agrees with other fluorites reported in the literature \cite{sheetal2022field,elghandour2024slow,ciomaga2016zirconate,clements2011fluorite}. Structural disorder is manifest in the site occupancies, i.e. mixing of \ce{\it{A}^{3+}}/\ce{\it{B}^{4+}} ions in the $4a$ site and 1/8 oxygen vacancy. Detailed results of positions and occupancies are shown in Table \ref{tab:table1}.

\begin{table}[h]
\caption{\justifying Positions and occupancies for \ce{Ho2Zr2O7}, refined from the data shown in Fig. \ref{fig:XRD}. The structural phase is $Fm\bar{3}m$ disordered-fluorite with lattice parameter $a=5.216(2)\text{\AA}$.}
\begin{ruledtabular}
\begin{tabular}{cccccc}
Atom& Site& x& y& z& Occ\\
\colrule
Ho & 4a & 0 & 0 & 0 & 0.50(1)\\
Zr & 4a & 0 & 0 & 0 & 0.49(1)\\
O & 8c & 0.25 & 0.25 & 0.25 & 0.87(1)\\
\end{tabular}
\end{ruledtabular}
\label{tab:table1}
\end{table}

\subsection{Inelastic Neutron Scattering}

\begin{figure}
    \centering
    \includegraphics[width=0.9\linewidth]{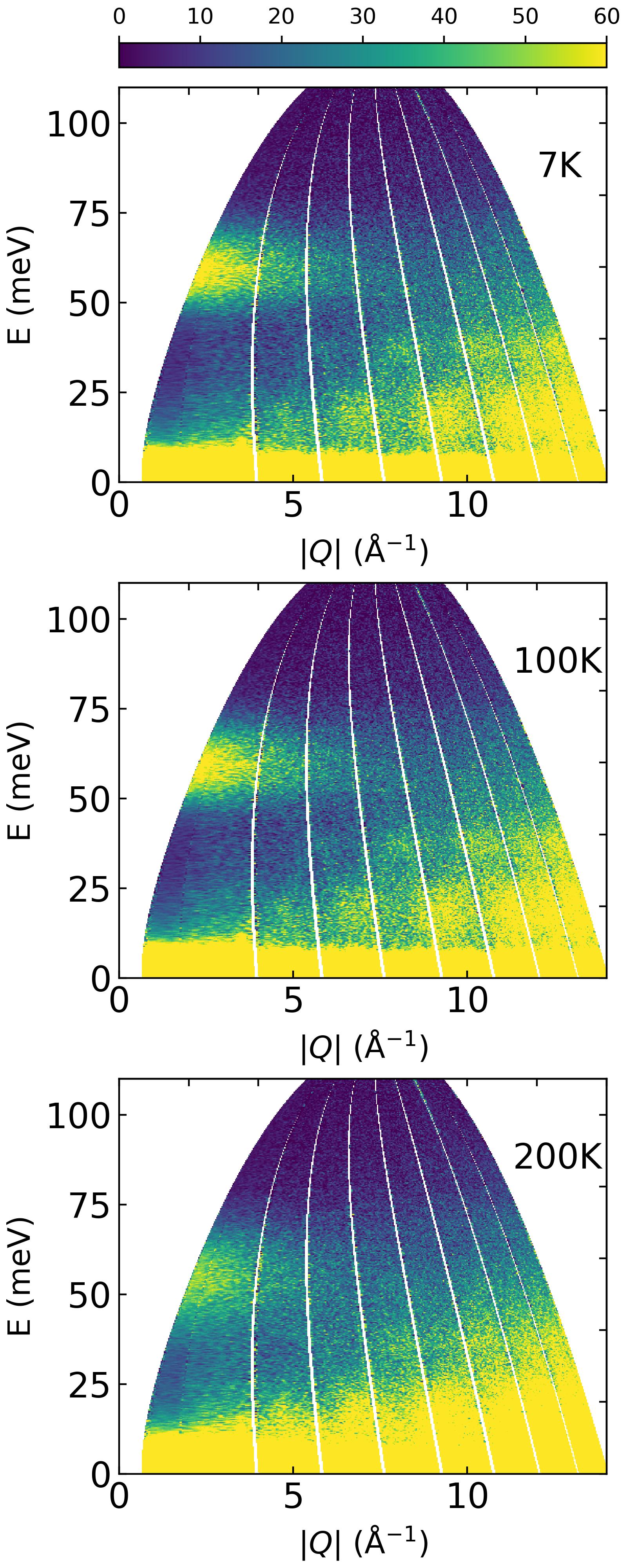}
    \caption{\justifying Inelastic neutron scattering measured with $E_\mathrm{i}=120$\,meV at 7\,K (top), 100\,K (middle) and 200\,K (bottom). A broad CEF excitation is observed around 60\,meV at all temperatures.}
    \label{fig:spectras}
\end{figure}

\begin{figure}
    \centering
    \includegraphics[width=\linewidth]{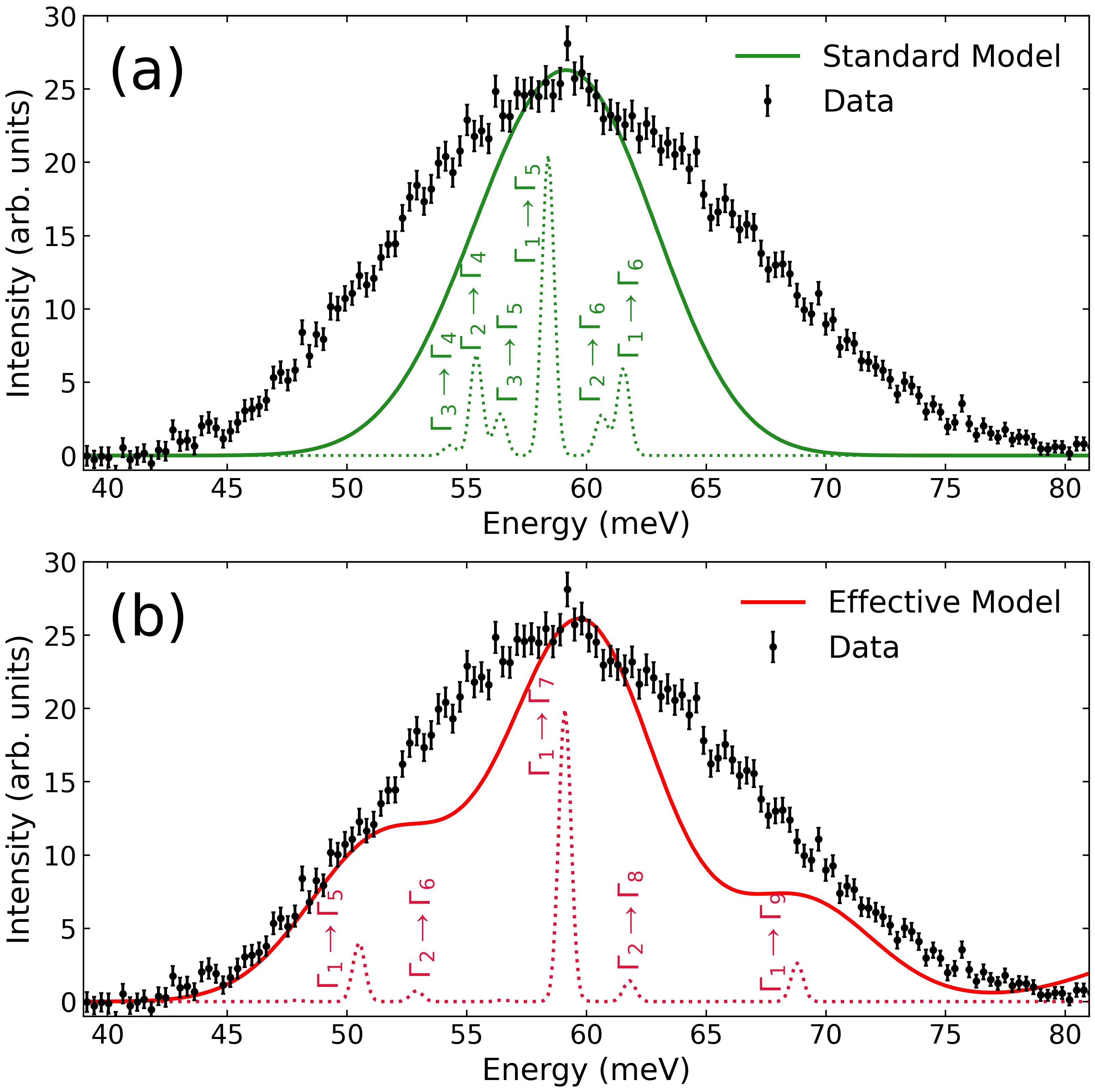}
    \caption{\justifying CEF excitation at around 60meV with the calculated neutron cross-section for each model at 7K. The dotted curves within the peak help to visualize the position of the transitions for (a) the standard model and (b) the effective model.
}
    \label{fig:cross-section}
\end{figure}

Data recorded on MARI at 7, 100 and 200\,K are shown in Fig. \ref{fig:spectras}. A broad scattering signal, ranging from about 40 up to 80\,meV, is measured at all temperatures. The intensity of this mode decreases with increasing momentum transfer $|\textbf{Q}|$ and temperature, confirming its magnetic origin. An additional peak near 2\,meV might be present in the $E_\mathrm{i}=10$\,meV data collected at 7\,K (see Appendix~\ref{appendixA}), however the experimental resolution does not allow its intensity to be separated from the elastic line. Figure \ref{fig:cross-section}(a)-(b) shows a cut along energy (black circles with error bars), performed integrating the signal in the interval $|\textbf{Q}|=[2:4]$ \AA$^{-1}$. Phonon scattering was assumed to contribute with a linear background, subtracted from the data shown in Fig. \ref{fig:cross-section}.  

In pyrochlores such as \ce{Ho2Ti2O7}, the CEF splits the spin-orbit ground-state manifold of the $J=8$ \ce{Ho^{3+}} ions into six doublets and five singlets \cite{rosenkranz2000crystal,ruminy2016crystal}. In \ce{Ho2Zr2O7}, on the other hand, the magnetic $4a$ site has a nominally higher, cubic $O_h$ point symmetry \cite{gorller1996rationalization,crystaltable}, for which the CEF Hamiltonian is written
\begin{equation}
\begin{split}
\mathcal{H}_\mathrm{CF}^\mathrm{S}&=B_0^4C_0^4+B_4^4(c)(C_{-4}^4+C_4^4)\\
&+B_0^6C_0^6+B_4^6(c)(C_{-4}^6+C_4^6),
\end{split}
\label{HfluoS}
\end{equation}
where $B_n^m$ and $C_n^m$ are the Wybourne parameters and tensors operators, respectively.
For perfectly cubic environments, only two crystal field parameters in Eq. (\ref{HfluoS}) are independent, as $B_4^4=0.5976B_0^4$ and $B_4^6=1.871B_0^6$ \cite{hutchings1964point}. 

The magnetic neutron scattering cross-section for a transition from a CEF level $\ket{\Gamma_j}$ to a level $\ket{\Gamma_i}$ is given by
\begin{equation}
\begin{split}
	S(\textbf{Q},\omega)\propto g_J^2\mu_\mathrm{B}^2f^2(\textbf{Q})\sum_i p_i \sum_j |\bra{\Gamma_i}\textbf{J}_\perp \ket{\Gamma_j}|^2 \\\times \delta(E_j-E_i-\hbar \omega),
\end{split}
    \label{crossneutron}
\end{equation}
where $g_J$ is the Landé g-factor, $\mu_\mathrm{B}$ is the Bohr magneton and $f(\mathbf{Q})$ is the magnetic form factor of the Ho$^{3+}$ ion in the dipole approximation. The factor $p_i$ is the population of the \emph{i}-th CEF level and $\mathbf{J}_\perp$ is the component of the total angular momentum operator perpendicular to the scattering vector $\mathbf{Q}$ \cite{boothroydneutron}. Initial data modeling was performed with the PyCrystalField package \cite{scheie2021pycrystalfield}, and the final fit was obtained with the software SPECTRE \cite{boothroydspectre}, when the cubic constraint to the parameters of the S model was enforced. The crystal-field parameters $B_n^m$ corresponding to the best fit of Eq.~(\ref{HfluoS}) to the data are listed in Table \ref{tab:table2}.

Calculated total (continuous line) and individual level intensities (dashed lines) for $\mathcal{H}_\mathrm{CF}^\mathrm{S}$ are shown in Fig.~\ref{fig:cross-section}(a). The instrumental resolution function is estimated from the elastic line to be approximately Gaussian with a full width at half maximum (FWHM) of $\sim7$\,meV. This value, which sets the maximum FWHM of the CEF levels for a time-of-flight experiment, was used in the calculation of the total intensities for both models. For clarity, the intensities of individual levels, as labeled in Tab.\ref{tab:table3}, were calculated with a fixed $\mathrm{FWHM}=0.1$\,meV. 

Clearly, the standard model is successful in predicting the median position of the measured excitations. Indeed, no other CEF levels are 
detected above $80$\,meV or down to $\sim3~\textrm{meV}$, where an unresolved mode can be inferred from higher-resolution data, as shown in Appendix~\ref{appendixA}. Importantly, however, the S model fails to account for the observed broadening of the level appearing around 60\,meV. As shown in Tab.~\ref{tab:table3} and Fig.~\ref{fig:cross-section}(a), although multiple levels are predicted around 60\,meV, they appear too close to the strongest excitation $\Gamma_1$→$\Gamma_{5}$. 

The effective (E) model is constructed as an effort to account for this broadening, while trying to capture the less symmetrical environments due to the oxygen vacancies. The Hamiltonian $\mathcal{H}_\mathrm{CF}^\mathrm{E}$ is an approximation considering arrangements with coordination numbers ($\mathrm{CN}$) 6 and 7. When $\mathrm{CN}=7$, the local symmetry around the \ce{Ho^{3+}} is the same, independent of the site of the oxygen vacancy. That is not true for $\mathrm{CN}=6$, in which different arrangements are allowed. In this case, after structure stability considerations, we assumed that the most probable configuration had two O$^{2-}$ vacancies at the cubic-diagonal axis (see Fig.~\ref{fig:oxygencn}). This choice preserves the 3-fold rotation around the local $\langle 111 \rangle$ axis, and the inversion point symmetry of the magnetic ion. 

Once the most probable local oxygen configurations were determined, we used the point-charge model \cite{hutchings1964point} to have a numerical estimate of each symmetry allowed $B_n^m$ for $\mathrm{CN}=6,7$ (see Appendix~\ref{appendixB} for more details). When adjusting the data shown in Fig.~\ref{fig:cross-section}, we gradually added the $B_n^m$ parameters which had the largest point-charge magnitude to the fit, making several attempts to adjust the data before adding another parameter. After this process, the final effective Hamiltonian was determined to be 
\begin{equation}
\begin{split}
\mathcal{H}_\mathrm{CF}^\mathrm{E}&=B_1^2(c)(C_{-1}^2-C_1^2)+iB_1^2(s)(C_{-1}^2+C_1^2)
\\&+iB_2^2(s)(C_{-2}^2-C_2^2)+B_0^4C_0^4+B_4^4(c)(C_{-4}+C_4^4)\\
&+B_0^6C_0^6+B_4^6(c)(C_{-4}^6+C_4^6).
\end{split}
\label{HfluoE}
\end{equation}
This model includes seven uncorrelated Wybourne parameters, all accounting for lower-symmetry configurations present in \ce{Ho2Zr2O7}. The crystal-field parameters $B_n^m$ corresponding to the best fits of Eq.~(\ref{HfluoE}) to the data are listed in Table \ref{tab:table2}. The calculated cross-section for the effective model is shown in Fig.~\ref{fig:cross-section}(b).There are now five spread excitations in the region between 40\,meV and 80\,meV, leading to a broader and asymmetric excitation mode. Compared to the standard model, there are no longer stronger transitions from the level $\Gamma_3$ near 2\,meV. However, the main excitation remains centered near 60\,meV, such as the experimentally observed peak.

The eigenvalues for Eqs. (\ref{HfluoS}) and (\ref{HfluoE}), along with level degeneracy, are shown in Table \ref{tab:table3}. In Table \ref{tab:table4}, ground-state wavefunctions for the S and E models, along with the first-excited state for $\mathcal{H}_\mathrm{CF}^\mathrm{E}$, are listed. The S and E models have a doublet and a singlet ground state, respectively. Interestingly, both $\Gamma_1$ wavefunctions have zero magnetic moment (see below), and both models present a narrow gap of $\sim0.9$\,meV to the first-excited state $\Gamma_2$. We have additionally calculated
the single-ion magnetization and the electronic specific heat
given by S and E models. The results are
discussed in the next section.

\begingroup
\renewcommand{\arraystretch}{1.25} 
\begin{table}[t]
\caption{\justifying Wybourne crystal field parameters calculated for the standard (S) and effective (E) models. The values are given in meV.}
\begin{ruledtabular}
\begin{tabular}{cccccccc}
Model & $B_0^4$& $B_4^4(c)$&$B_0^6$&$B_4^6(c)$&$B_1^2(c)$&$B_1^2(s)$& $B_2^2(s)$\\ 
\hline
S&-333.8& -199.5& 68.7& -128.6&-&-&-\\ 
E&-186.1& -84.8& 185&-134&-64.3& -64.3&-347    
\end{tabular}
\end{ruledtabular}
\label{tab:table2}
\end{table}
\endgroup

\begingroup
\renewcommand{\arraystretch}{1.25} 
\begin{table}[b]
\caption{\justifying Calculated CEF eigenstates and their degeneracies in the two different point-charge models. Model level degeneracies are stated in parentheses.}
\begin{ruledtabular}
\begin{tabular}{cccc}
 \multicolumn{2}{c}{Standard} & \multicolumn{2}{c}{Effective} \\
 \hline
 Level & Energy (meV) & Level & Energy (meV)\\ 
 \hline
$\Gamma_1$ (2) &0.000& $\Gamma_1$ (1) & 0.000 \\
$\Gamma_2$ (3) &0.866& $\Gamma_2$ (1) &0.938\\
$\Gamma_3$ (3) &2.035& $\Gamma_3$ (1) &2.591 \\
$\Gamma_4$ (2)&56.311&$\Gamma_4$ (1)&3.711\\
$\Gamma_5$ (3)&58.353&$\Gamma_5$ (1)&50.532\\
$\Gamma_6$ (3)&61.550&$\Gamma_6$ (1)&53.791\\
$\Gamma_7$ (1)&67.716&$\Gamma_7$ (1)&59.125\\
  & &$\Gamma_8$ (1)&62.723\\
  & &$\Gamma_9$ (1)&68.777\\
  & &$\Gamma_{10}$ (1)& 76.558\\
  & &$\Gamma_{11}$ (1) &84.028\\
  & &$\Gamma_{12}$ (1)&85.486\\ 
  & &$\Gamma_{13}$ (1)&87.065\\
  & &$\Gamma_{14}$ (1) &108.489\\
  & &$\Gamma_{15}$ (1)&113.042\\
  & &$\Gamma_{16}$ (1)&143.509\\
  & &$\Gamma_{17}$ (1)&144.220\\
\end{tabular}
\end{ruledtabular}
\label{tab:table3} 
\end{table}
\endgroup

\begingroup
\renewcommand{\arraystretch}{1.25} 
\begin{table*}
\caption{\label{tab:table4} Calculated wavefunctions of the ground ($\Gamma_1$) and first excited ($\Gamma_2$) states for the standard (S) and effective (E) models. Angular momentum components $\ket{m_j}$ with coefficients $<0.07$ are omitted from the wavefunctions for clarity. 
}
\begin{ruledtabular}
\begin{tabular}{ccccc}
 Level & \multicolumn{4}{c}{Wavefunction} \\ 
 \hline
 & \multicolumn{4}{c}{Standard model} \\
 \hline
$\Gamma_1$&\multicolumn{4}{c}{$\mp0.255\ket{\pm8}+0.591\ket{\pm6}\pm0.645\ket{\pm4}-0.388\ket{\pm2}\pm0.194\ket{0}$} \\ 
$\Gamma_2$&\multicolumn{4}{c}
{$\ket{\Gamma_2^1}=0.6(\ket{+6}+\ket{-6})+0.533(\ket{-2}+\ket{+2})$}\\
&\multicolumn{4}{c} {$\ket{\Gamma_2^{2}}=-0.149\ket{7}+0.175\ket{5}+0.897\ket{-5}+0.195\ket{3}-0.314\ket{-1}$}\\
&\multicolumn{4}{c}{$\ket{\Gamma_{2}^{3}}=-0.149\ket{-7}+0.897\ket{5}-0.175\ket{-5}+0.195\ket{-3}-0.314\ket{1}$} \\
\hline
& \multicolumn{4}{c}{Effective model} \\
\hline
$\Gamma_1$&\multicolumn{4}{c}{$0.209(\ket{-8}-\ket{+8})+0.274i(\ket{+6}+\ket{-6})+0.518(\ket{+4}-\ket{-4})-0.320i(\ket{+2}-\ket{-2})$}\\
$\Gamma_2$&\multicolumn{4}{c}{$-0.210(\ket{+8}-\ket{-8})+0.271i(\ket{+6}-\ket{-6})+0.517(\ket{+4}+\ket{-4})+0.302i(\ket{-2}-\ket{+2})-0.185\ket{0}$}

\end{tabular}
\end{ruledtabular}
\end{table*}
\endgroup

\subsection{Thermomagnetic characterization}

\begin{figure}
    \centering
    \includegraphics[width=\linewidth]{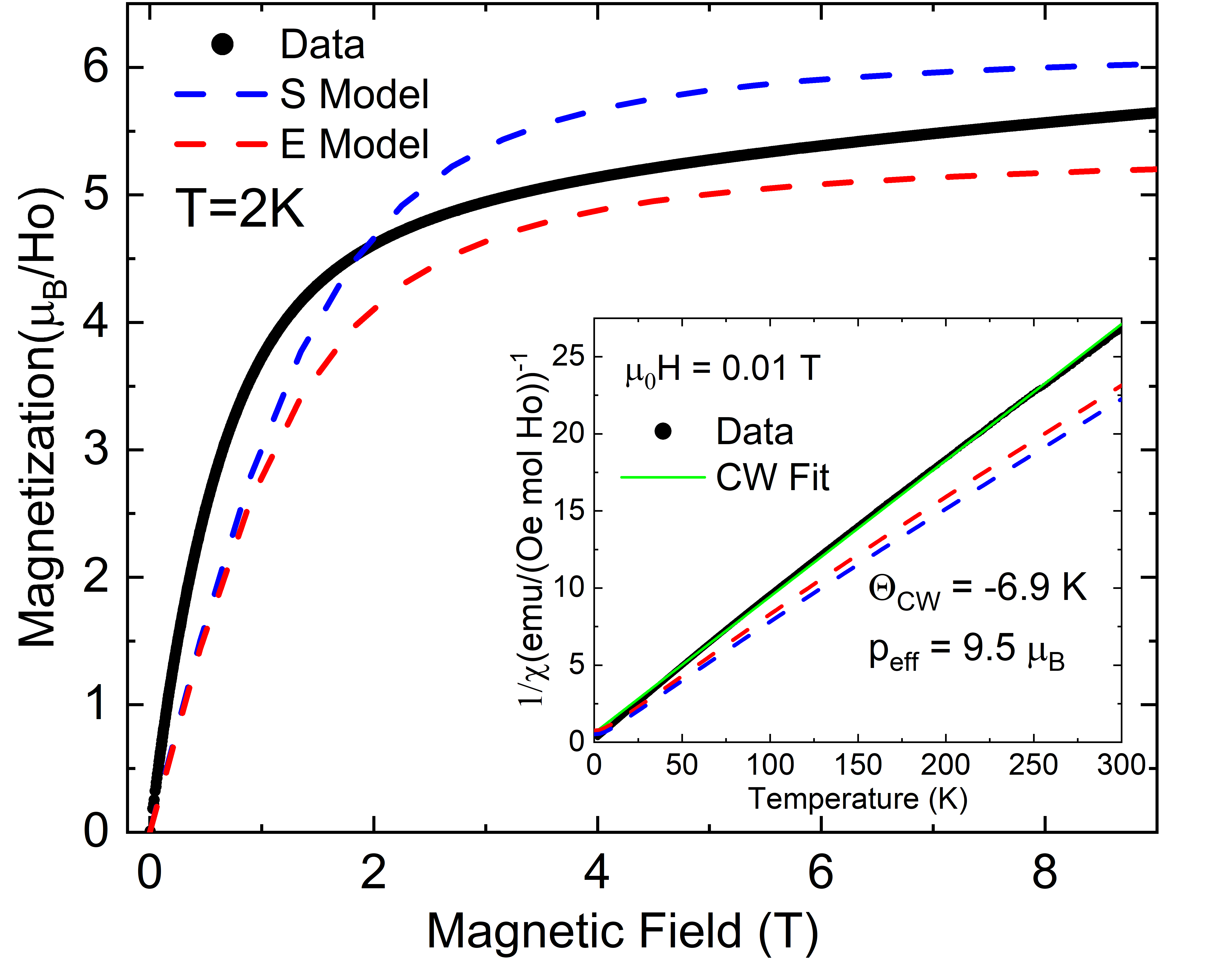}
    \caption{\justifying Magnetization versus the applied magnetic field at 2K. The dashed lines are the single-ion magnetizations calculated using the CEF steven parameters obtained by the standard (S) and effective (E) point-charge models. Inset:  temperature dependence of the inverse magnetic susceptibility, the Curie-Weiss temperature, and effective magnetic moment determined by the Curie-Weiss fitting (green line).}
    \label{fig:magnetization}
\end{figure}
The DC magnetization of \ce{Ho2Zr2O7}, measured at 2\,K as a function of magnetic fields up to 9\,T, is shown in Fig.~\ref{fig:magnetization}). The measured curve does not fully saturate, but for higher fields remains close to $5\,\mu_\mathrm{B}$/ion, half of the value for the \ce{Ho^{3+}} free ion. This behavior is also present in other Ising pyrochlores \cite{fukazawa2002magnetic,lhotel2012low}, including \ce{Ho2Ti2O7} \cite{krey2012first}, and is consistent with the literature on the compound \cite{elghandour2024slow}. From a linear fitting of the inverse DC magnetic susceptibility (see inset in Figure \ref{fig:magnetization}), we obtain a Curie-Weiss temperature of $-6.9$\,K and an effective magnetic moment of $9.5\,\mu_\mathrm{B}$. The effective moment is close to the \ce{Ho^{3+}} free-ion value of $10.6\,\mu_\mathrm{B}$. The effective coupling  $\mathcal{J}_\mathrm{eff}$  for the nearest neighbors can be estimated using a mean-field approximation expression $\mathcal{J}_\mathrm{eff}=3\theta_\mathrm{CW} /zJ(J+1)$, where $z$ is the number of first neighbors. For \ce{Ho2Zr2O7}, $z=3$ and $J=8$, resulting in $\mathcal{J}_\mathrm{eff}=0.1$\,K.  

\begin{figure}[b]
    \centering
    \includegraphics[width=\linewidth]{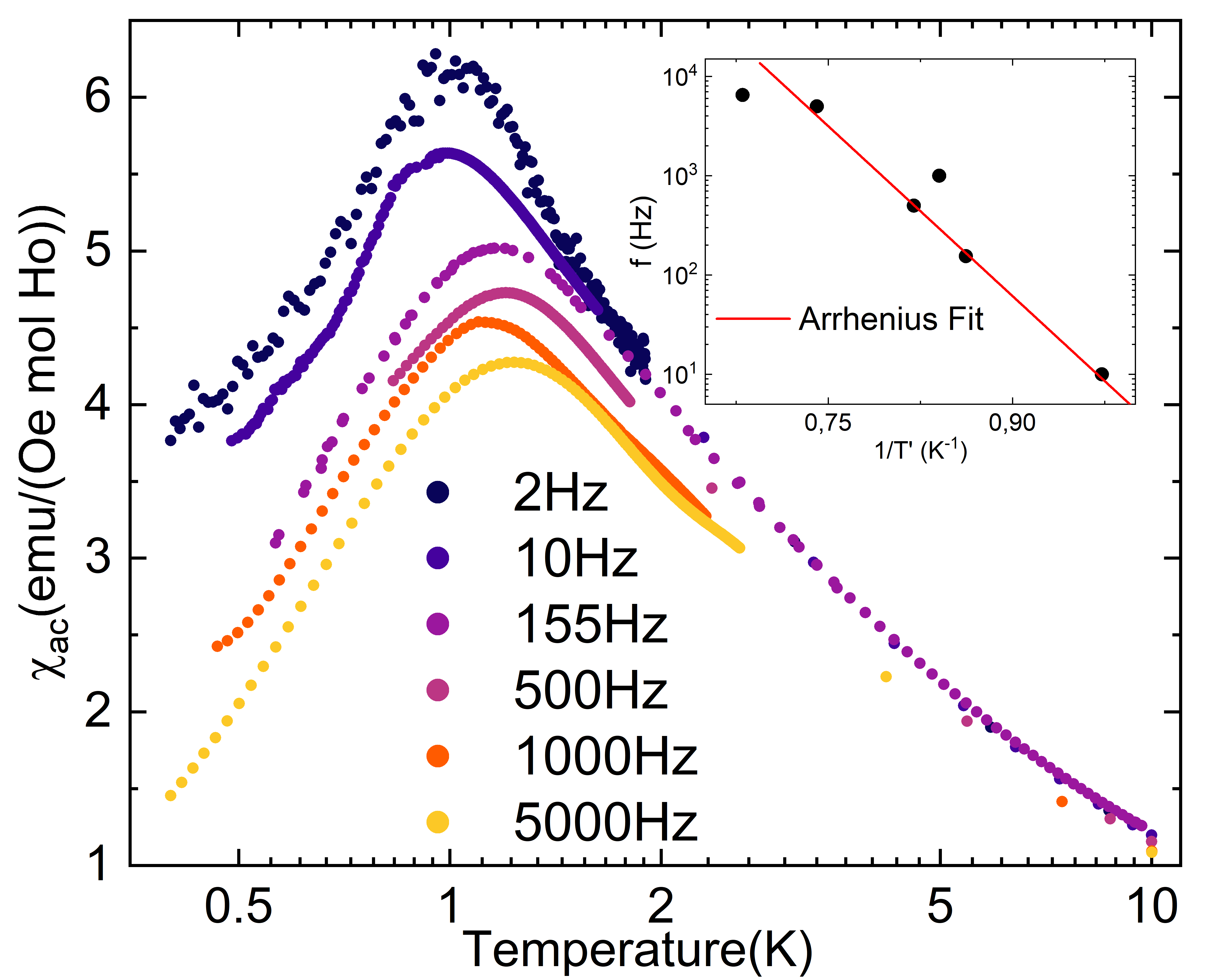}
    \caption{\justifying Temperature dependence of the real part of the AC Magnetic Susceptibility. Inset: dependence of the frequency versus the inverse of the temperature of the maximum in the magnetic susceptibility, with the Arrhenius law fit.}
    \label{fig:xac}
\end{figure}

AC magnetic susceptibility is used to investigate spin dynamics at low temperatures. The real part of these measurements down to $0.5$\,K are shown in Fig.~\ref{fig:xac}. The response is paramagnetic down to $T^{*} = 1$\,K, when a frequency-dependent maximum occurs. The data are well described by the Arrhenius law $f=f_0\exp(-E_\mathrm{b}/k_\mathrm{B}T)$, where $E_\mathrm{b}$ is the energy barrier, $f_0$ the characteristic frequency, and $T$ is the temperature of the maximum. This fit is shown in the inset of Fig.~\ref{fig:xac}. We estimated $E_\mathrm{b}=26$\,K, a relaxation time $\tau_0=f_0^{-1}=8.10^{-13}$\,s, and a temperature shift of the maximum per decade frequency of $\delta T'=0.11$ defined as $\delta T'=\Delta T'/T'\Delta \log (f)$.

\begin{figure}
    \centering
    \includegraphics[width=\linewidth]{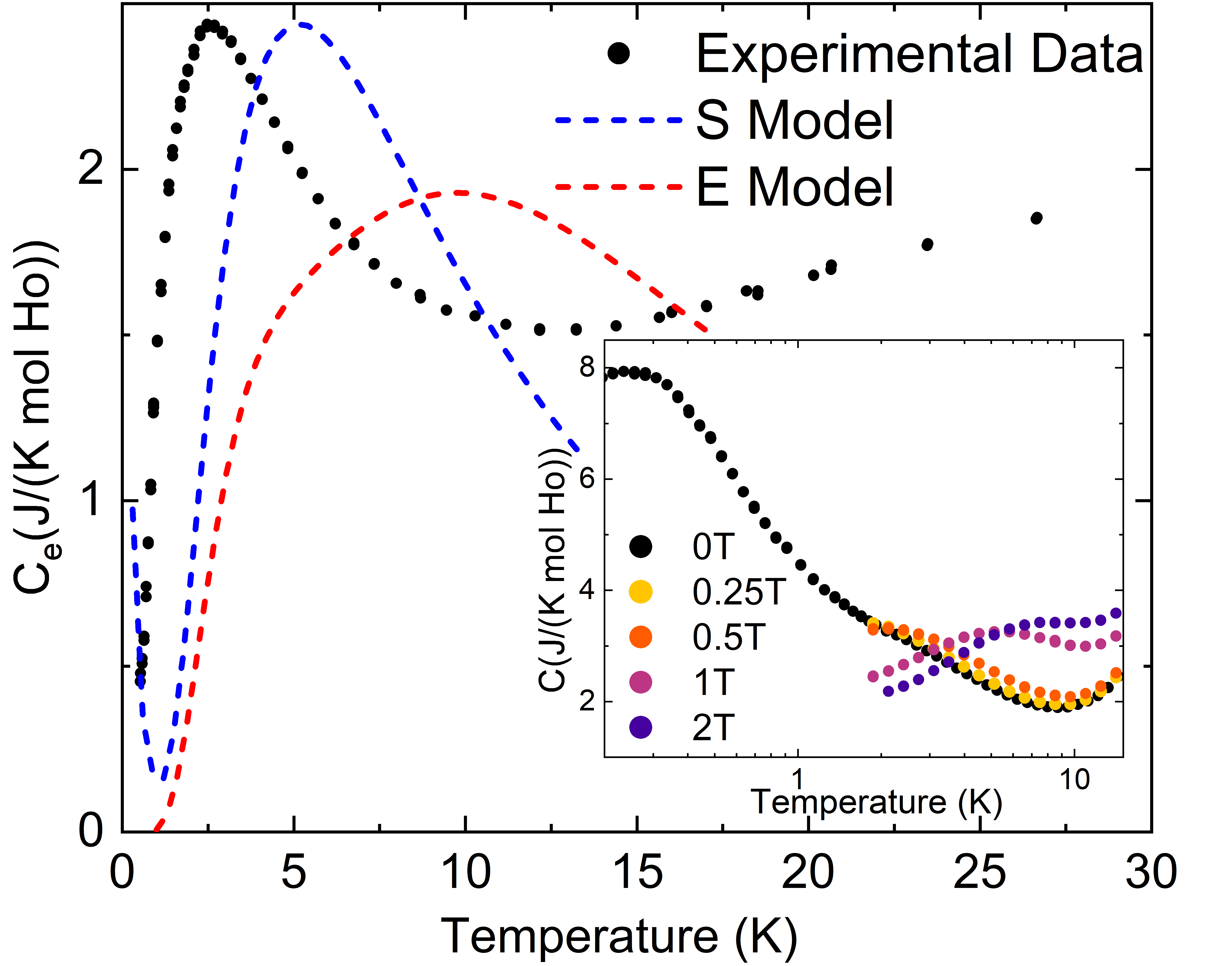}
    \caption{\justifying The electronic specific heat versus the temperature in zero magnetic field. The dashed curves were calculated using the CEF parameters for the standard (S) and effective (E) point-charge models. Inset: total specific heat in different magnetic fields. No sign of long-range order was found down to $200$\,mK.}
    \label{fig:specificheat}
\end{figure}

Electronic magnetic specific heat $C_\textrm{e}$, measured down to $150$\,mK, is displayed in Figure \ref{fig:specificheat}. The nonmagnetic fluorite \ce{Lu2Zr2O7} was used to estimate the phonon contribution to the total specific heat, which was subtracted from the data. The low-temperature nuclear specific heat was accounted for by considering the hyperfine and quadrupolar Hamiltonian for the single isotope \ce{^{165}Ho} ($I = 7/2$) \cite{bleaney,lounasmaaholmium}. As shown in the inset of Fig.~\ref{fig:specificheat}, no sign of long-range order is found. The shoulder observed at $T=2$\,K in \ce{Ho2Ti2O7} \cite{lau2006zero} is also present in \ce{Ho2Zr2O7}, but appears broader and shifted to a higher temperature ($T=3$\,K). The inset of Fig.~\ref{fig:specificheat} shows that this shoulder is suppressed by magnetic field.

We tried to fit the electronic specific heat using a two-level Schottky model, estimating the gaps between the states. As expected based on our measured CEF level-scheme, the Schottky model does not reproduce the broad feature in $C_\textrm{e}$ satisfactorily, although the gaps showed a linear dependence with the field (not shown). Therefore, using the Zeeman splitting for a $S=1/2$ spin, $E=(g\mu_\mathrm{B}/k_\mathrm{B})H$ we estimate a g-factor of $6.8$, smaller than the $g_ {||}=19.6$ along the $\langle 111 \rangle$ axis reported for \ce{Ho2Ti2O7} \cite{bertin2012crystal}. The entropy (not shown) of \ce{Ho2Zr2O7} neither saturates around $15$~K nor at the residual entropy of the spin ice $(R/2)\ln(3/2)$ unlike observed in \ce{Ho2Ti2O7} or \ce{Dy2Ti2O7} \cite{ramirez1999zero}. 

\section{DISCUSSION}

In \ce{Ho2Ti2O7}, the first excited state is at around $20$\,meV\cite{bertin2012crystal}. The main magnetic feature in our inelastic neutron scattering data is the significant spectral weight centered around 60 meV, with a width remarkably larger than the instrumental resolution. A similar situation occurred in Porée et al. \cite{poree2022crystal}, where the authors attributed the broadened peaks in \ce{Ce2Hf2O7} to impurities, adjusting empirically the observed FWHM of the modes in their model. In fact, some of our experimental data can be understood with the S model, Eq. (\ref{HfluoS}), which successfully predicts the average energy of the measured CEF levels, and gives a good estimate of the compound's magnetization and electronic specific heat. However, the calculated intensities for the transitions of the S model at $7$\,K [Fig.~\ref{fig:cross-section}(a)] are too close to each other to account for the total width of the observed excitations. The inclusion of additional terms in the Hamiltonian of the E model in Eq. (\ref{HfluoE}) led to an increased separation of the high-energy excitations, while keeping the most intense transition centered at 60\,meV.

The half-saturation behavior observed in the magnetization of titanate pyrochlores is attributed to strong single-ion anisotropy \cite{bramwell2000bulk} and to the geometry of the magnetic sublattice. Given the anion disorder present in this system, the magnetic properties calculated with both E and S models agree remarkably well with the measured bulk data. They reproduce not only the saturation trend in Fig.~ \ref{fig:magnetization}, but also the measured slope in the inverse susceptibility. The small discrepancy at $2$\,K may stem from magnetic interactions, particularly dipolar-dipolar, given the relatively large effective moment at this temperature, spin-lattice coupling and disorder. 

The height and width of the AC peak resemble those measured in the canonical spin ices \ce{Ho2Ti2O7} and \ce{Dy2Ti2O7} \cite{quilliam2011dynamics,matsuhira2002new}, as well other zirconate defect-fluorites \ce{Dy2Zr2O7} and \ce{Tb2Zr2O7} \cite{ramon2019absence,ramon2023glassy}. The reported Arrhenius fit values for \ce{Ho2Ti2O7} in the spin ice regime are $E_\mathrm{b} = 28$\,K and $\tau_0 = 2.10^{-14} $\,s \cite{quilliam2011dynamics,matsuhira2000low}, both close to our results. The energy scale of $E_\mathrm{b}$ is close to the $\Gamma_{3}$ CEF level calculated in the S and E model, and the peak in AC susceptibility may be a signature the depopulation of this particular state. The peak temperature shift with frequency is an order of magnitude larger than that of the canonical spin glasses ($0.005-0.01$) \cite{mydosh1993spin}. That, since the AC magnetic susceptibility decreases but does not vanish down to $0.5$\,K, indicates the setting of an antiferromagnetically coupled frustrated ground state. This aligns with the conclusions of Elghandour \emph{et al.} \cite{elghandour2024slow}, which characterized a frozen spin state below $0.6$\,K in \ce{Ho2Zr2O7}.

Similarly to \ce{Ho2Ti2O7} \cite{lau2006zero}, no signature of long-range order was observed in \ce{Ho2Zr2O7} specific heat. Our CEF analysis, in addition to reproducing fairly well its behavior, provides an explanation for the difficulty in fitting a two-level Schottky anomaly to $C_\mathrm{e}(T)$. Notably, the ground-state g-factors $g_{||}=2g_J|\bra{\Gamma_1^{\pm}}\textbf{J}_z \ket{\Gamma_1^{\pm}}|$ and $g_{\perp}=g_J|\bra{\Gamma_1^{\pm}}\textbf{J}_\pm \ket{\Gamma_1^{\mp}}|$ \cite{bertin2015geometrical} of S and E models are zero, even though $\Gamma_1$ for $\mathcal{H}_\mathrm{CF}^\mathrm{S}$ is a (non-Kramers') doublet. The models also agree with one another on the energy of the first excited state(s), of only $\sim0.9$\,meV. These results demonstrate that a simple effective spin-$\tfrac{1}{2}$ Ising ground state is not present in \ce{Ho2Zr2O7}. Instead, the magnetism observed in the fluorite is a direct consequence of the small magnitude of the CEF gap tuned by structural disorder, which allows for the mixing of levels at finite temperatures. 

\section{CONCLUSION}

In conclusion, our DC magnetic susceptibility measurements performed on \ce{Ho2Zr2O7} reveal local antiferromagnetic correlations between the Ho$^{3+}$ ions. The appearance of a broad peak in AC susceptibility at $T^*=1$\,K \cite{elghandour2024slow}, which also shows a weak frequency dependence, is consistent with slow spin dynamics at finite temperatures. No evidence of long-range magnetic ordering is observed at $T^*$ in specific heat measurements.

The CEF spectrum for \ce{Ho2Zr2O7} differs significantly from that of \ce{Ho2Ti2O7} in two aspects which are fundamental for the formation of the spin-ice state in the titanate. Firstly, our neutron and bulk measurements strongly suggest that the ground state of the fluorite, which has zero magnetic moment for both our CEF models, is not well isolated as in the pyrochlore, but separated by less than $1$\,meV from the first excited state. Secondly, the calculated ground-state wavefunctions for the zirconate have a significantly different level composition from the almost pure $\ket{m_j=\pm8}$ of the titanate. The occupation of these accessible excited levels may explain the persistent magnetic response we observe down to $200$\,mK. The presence of many accessible states may enhance quantum fluctuations, an important feature for the formation of exotic magnetic states \cite{gingras1,poree2022crystal}.
 
Incorporating additional CEF parameters to model disordered systems is a non-trivial task, especially in compounds containing non-Kramers ions with large total angular momentum $J$. We believe the modeling approach to disorder in this work may assist future CEF analysis of pyrochlores, defect-fluorites, and other disordered compounds.

\section*{ACKNOWLEDGMENTS}

P. L. Oliveira Silva and V. Peçanha-Antonio thank Prof. A. T. Boothroyd for discussions and inputs, fundamental to the CEF analysis presented in this work. P. L. O. Silva thanks the financial support of CAPES (Grant No. 88887.835016/2023-00) and FAPESP (Grant No. 2023/12679-0). Some of the research (J. S. Gardner) was supported by the U.S. Department of Energy, Office of Science, Basic Energy Sciences. The ORNL is managed by UT-Battelle, LLC, under Contract No.DE-AC05-00OR22725. R. S. Freitas acknowledges FAPESP (Grant No. 2021/12470-8) and CNPq (Grant No. 312341/2021-0). Experiments at the ISIS Neutron and Muon Source were supported by beamtime allocation RB2010678 from the Science and Technology Facilities Council. The raw data used in this work can be found at \cite{neutrondata}.

\appendix

\section{LOW-ENERGY CEF EXCITATION}\label{appendixA}

\begin{figure}[h]
    \centering
    \includegraphics[width=\linewidth]{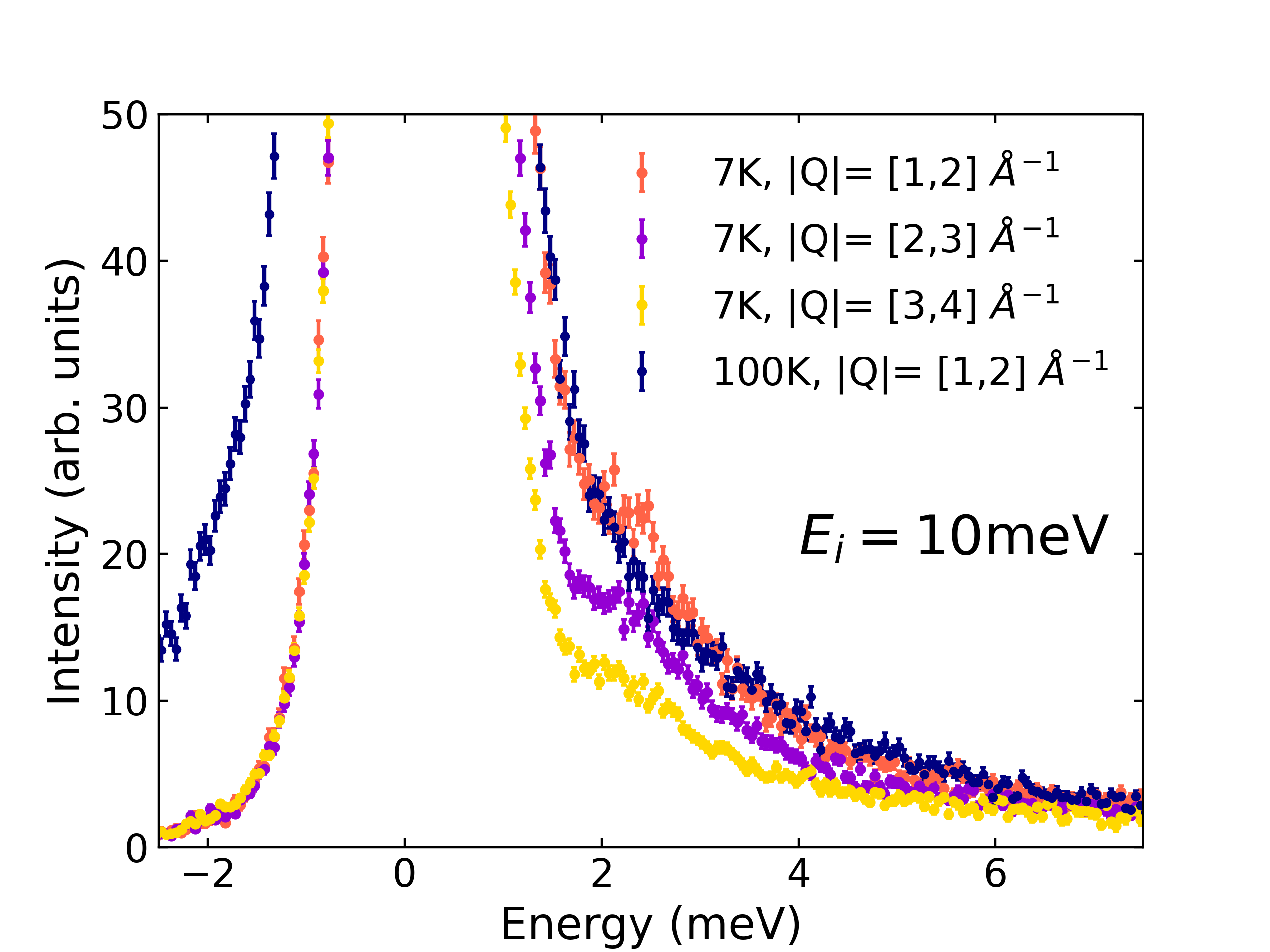}
    \caption{\justifying CEF excitation near the elastic line for the data of $E_i=10$\,meV at $7$\,K and $100$\,K. There is not enough resolution to resolve the peaks in this region.}
    \label{fig:10meV}
\end{figure}

In Fig. \ref{fig:10meV}, cuts along energy for the data recorded at $7$\,K and $100$\,K with an incident energy of $E_i=10$\,meV are shown. At $7$\,K, data were integrated over three distinct intervals of momentum transfer, $|\textbf{Q}|=[1:2]$ \AA$^{-1}$, $|\textbf{Q}|=[2:3]$ \AA$^{-1}$, and $|\textbf{Q}|=[3:4]$ \AA$^{-1}$, while at $100$\,K, only the first interval is shown for clarity. The intensity of the signal at $\sim 2$~meV decreases with increasing $|\textbf{Q}|$ and temperature, characteristics consistent with what is expected from a CEF mode. Our models predict the presence of this level, and another one at even lower energies ($\sim0.9$~meV), but both seem to overestimate the intensity of the excitations in this region.

\section{EFFECTIVE POINT-CHARGE MODEL IN \ce{Ho2Zr2O7}}\label{appendixB}

The first step in constructing the effective point-charge model was to determine the prevalence of oxygen-deficient environments around \ce{Ho^{3+}}. Since the magnetic ion is shared by eight unit cells, the probability of having a coordination number N is given by a binomial distribution

\begin{equation}
    P_\mathrm{CN}(N)=\frac{8!}{N!(8-N)!}\left(\frac{7}{8}\right)^N\left(\frac{1}{8}\right)^{8-N}.
    \label{eqn_appendix}
\end{equation} 
The probabilities for 6, 7 and 8 O$^{2-}$ neighbors are 19.4\%, 39.3\%, and 34.4\%, respectively, accounting for over 93\% of cases. We therefore focused our analysis on, apart from the standard cubic environment, the two additional cases with $\mathrm{CN} = 6$ and $\mathrm{CN} = 7$. 

\begin{figure}[h]
	\centering
	\includegraphics[width=0.5\linewidth]{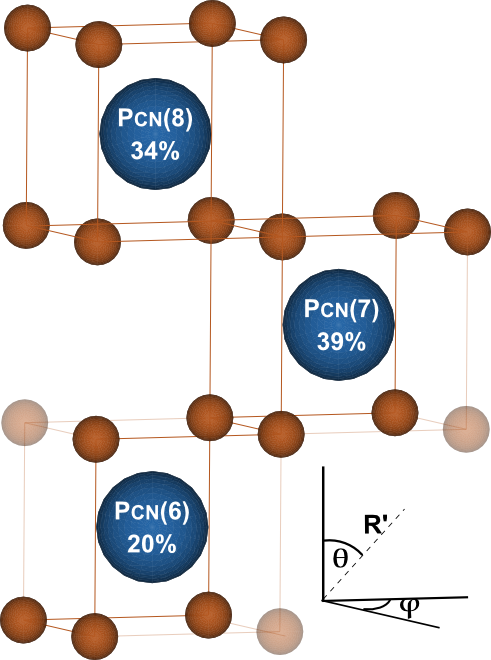}
	\caption{\justifying Oxygen configurations considered to the effective point-charge model in the defect-fluorite. The orange spheres are the oxygens and the blue ones are the holmium ions. Next to them there is the probability associated with each of the coordination numbers (CN).}
	\label{fig:oxygencn}
\end{figure}

Next, we calculated the non-zero Stevens parameters for each case using the point charge model. As briefly explained in the main text, we considered the local environments as illustrated in Fig. \ref{fig:oxygencn}. The origin of our coordinate system is fixed at the rare-earth atom. In the case of $\mathrm{CN} = 7$, the symmetry of the final arrangement is the same independently of the vacant oxygen. However, for $\mathrm{CN} = 6$, among the many possibilities, we selected the one that preserves $\bar{3}$ point symmetry of the rare-earth.

For analytical calculations of the point-charge model, we followed Hutchings’ original point-charge model methodology \cite{hutchings1964point}. The coordinate system used to define the oxygen positions for each CN is the same, and the axis used are defined in Fig. \ref{fig:oxygencn}. Table \ref{tab:coordinates} shows the O coordinates for each case. The general form for the Steven's parameters is $B_m^n=f_m^nZe^2\frac{\expval{r^m}}{R^{m+1}}$, where $f_m^n$ is a numeric factor, $\expval{r^m}$ is the expectation value of the n-th power of the distance electrons-nucleus for the $4f$ electrons, and R is the distance between Ho and O \cite{hutchings1964point}. To compute them, we used the multiplicative factors in references \cite{hutchings1964point,edvardsson1998role}.

The third step was selecting which parameters to include in the model. The priority of inclusion was gauged by the theoretical point-charge values of these CEF parameters. After several attempts of fitting, we chose to retain only the three largest terms: $B_2^2(s)$, $B_1^2(s)$ and $B_1^2(c)$. These parameters were included in our effective model with the constraint $B_2^1(s)=B_2^1(c)$. Other parameters were neglected as their theoretical values were less than 3\% of $B_2^2(s)$. 

\begingroup
\renewcommand{\arraystretch}{1.25} 
\begin{table}[t]
\caption{\justifying Spherical coordinates $(R,\theta,\phi)$ of the oxygens for each coordination number (CN) to calculate the Steven parameters for the effective model. Here,  $\theta_1=\arctan(\sqrt2)$ and $R=\frac{\sqrt{3}}{4}a$, where \textit{a} is the lattice parameter.}
\begin{ruledtabular}
\begin{tabular}{ccccc}
CN & \multicolumn{4}{c}{Coordinates}\\ \hline
\multirow{3}{*}{8}&\multicolumn{4}{c}{$(R,\pi-\theta_1,\frac{\pi}{4}),(R,\pi-\theta_1,\frac{3\pi}{4}),(R,\pi-\theta_1,\frac{5\pi}{4})$}\\ 
&\multicolumn{4}{c}{$(R,\pi-\theta_1,\frac{7\pi}{4}),(R,\theta_1,\frac{\pi}{4}),(R,\theta_1,\frac{3\pi}{4})$}\\
&\multicolumn{4}{c}{$(R,\theta_1,\frac{5\pi}{4}),(R,\theta_1,\frac{7\pi}{4})$}\\ \hline 
\multirow{3}{*}{7}&\multicolumn{4}{c}{$(R,\pi-\theta_1,\frac{\pi}{4}),(R,\pi-\theta_1,\frac{3\pi}{4}),(R,\pi-\theta_1,\frac{5\pi}{4})$}\\
&\multicolumn{4}{c}{$(R,\theta_1,\frac{\pi}{4}),(R,\theta_1,\frac{3\pi}{4}),(R,\theta_1,\frac{5\pi}{4})$}\\
&\multicolumn{4}{c}{$(R,\theta_1,\frac{7\pi}{4})$}\\ \hline 
\multirow{2}{*}{6}&\multicolumn{4}{c}{$(R,\pi-\theta_1,\frac{\pi}{4}),(R,\pi-\theta_1,\frac{3\pi}{4}),(R,\pi-\theta_1,\frac{7\pi}{4})$}\\
&\multicolumn{4}{c}{$(R,\theta_1,\frac{3\pi}{4}),(R,\theta_1,\frac{5\pi}{4}),(R,\theta_1,\frac{7\pi}{4})$}

\end{tabular}
\end{ruledtabular}
\label{tab:coordinates}
\end{table}
\endgroup

\FloatBarrier

\bibliography{ref}

\end{document}